\documentclass[hidelinks,onefignum,onetabnum]{siamart220329}

\usepackage{epstopdf}
\usepackage[utf8]{inputenc} % allow utf-8 input
\usepackage[T1]{fontenc}    % use 8-bit T1 fonts
\usepackage{url}            % simple URL typesetting
\usepackage{booktabs}       % professional-quality tables
\usepackage{nicefrac}       % compact symbols for 1/2, etc.
\usepackage{microtype}      % microtypography
\usepackage{amssymb,amsmath,amsfonts,bbm}
\usepackage{changepage}
\usepackage{color,graphicx,float}
\usepackage{bm,enumerate}
\usepackage{alltt,listings,tikz}
\usepackage[all]{xy}
\usepackage{hyperref,cite}
\usepackage{dsfont}
\usepgflibrary{shadings}
\usepackage{algorithm,algpseudocode}

\usepackage[english]{babel}
\usepackage{enumitem}
\setlist[enumerate]{leftmargin=0.2in}
\setlist[itemize]{leftmargin=0.2in}

\newcommand{\ftheta}{f_{\theta}}
\newcommand{\gtheta}{g_{\theta}}
\newcommand{\ft}{\widetilde{f}}
\newcommand{\thetat}{\widetilde{\theta}}

\newcommand{\Ltheta}{\mathcal{L}_{\theta}}
\newcommand{\Stheta}{\mathcal{S}_{\theta}}

\newcommand{\Mtheta}{M_{\theta}}

\newcommand{\add}[1]{{\color{black}{#1}}}

\newsiamremark{remark}{Remark}
\newsiamremark{hypothesis}{Hypothesis}
\crefname{hypothesis}{Hypothesis}{Hypotheses}
\newsiamthm{claim}{Claim}

\ifpdf
  \DeclareGraphicsExtensions{.pdf,.pdf,.png,.jpg}
\else
  \DeclareGraphicsExtensions{.pdf}
\fi

\headers{Inference of interaction kernels in opinion models}{W. Chu, Q. Li, and M. A. Porter}

\title{Inference of interaction kernels in mean-field models of opinion dynamics}

\author{Weiqi Chu\thanks{Department of Mathematics, University of California, Los Angeles 
  }
\and Qin Li\thanks{Department of Mathematics, University of Wisconsin, Madison}
\and Mason A. Porter\thanks{Department of Mathematics, University of California, Los Angeles; Department of Sociology, University of California, Los Angeles; and Santa Fe Institute}
}
\usepackage{amsopn}

\begin{document}

\maketitle

% REQUIRED
\begin{abstract}
In models of opinion dynamics, many parameters---either in the form of constants or in the form of functions---play a critical role in describing, calibrating, and forecasting how opinions change with time. When examining a model of opinion dynamics, it is beneficial to infer its parameters using empirical data. In this paper, we study an example of such an inference problem. We consider a mean-field bounded-confidence model with an unknown interaction kernel between individuals. This interaction kernel encodes how individuals with different opinions interact and affect each other's opinions. Because it is often difficult to quantitatively measure opinions as empirical data from observations or experiments, we assume that the available data takes the form of partial observations of a cumulative distribution function of opinions. We prove that certain measurements guarantee a precise and unique inference of the interaction kernel and propose a numerical method to reconstruct an interaction kernel from a limited number of data points. Our numerical results suggest that the error of the inferred interaction kernel decays exponentially as we strategically enlarge the data set.
\end{abstract}

% REQUIRED

% REQUIRED

%%%%%

%%%%

\section{Introduction} \label{intro}

The opinions and associated actions of the individuals in a population have major effects on financial markets \cite{may2008ecology}, pandemic responses \cite{bavel2020using}, climate change \cite{otto2020social}, and many other phenomena \cite{Bak-Colemane2025764118}. There are many investigations of opinion dynamics in both applied and theoretical contexts \cite{galesic2021}, and the mathematical modeling of opinions is prevalent in many disciplines, including sociology, economics, political science, mathematics, and physics \cite{castellano2009statistical,bonacich2012,noor2020}. 

Models of opinion dynamics, the spread of information, and other social phenomena can take the form of either deterministic or stochastic processes on networks \cite{porter2016dynamical}. In such models, the nodes of a network encode social entities (such as individual people) and the edges of a network encode interactions between entities \cite{newman2018}. In a model of opinion dynamics, each entity of a social network holds a time-dependent opinion. Such opinions are continuous-valued in some models and discrete-valued in others~\cite{noor2020}. When entities interact with each other, they may adjust their opinions according to some update rule. The update rule and network structure jointly affect the steady-states and transient dynamics of opinion models. Relevant phenomena include how long it takes a system to converge to a steady state (and whether or not it does so) \cite{meng2018opinion}, whether or not the entities of a system eventually reach a consensus state \cite{noor2020}, and how extremist opinions can take root in a system and affect the overall dynamics~\cite{amblard2004role,ojer2022}.

Many studies of opinion models take a ``forward'' perspective and explore how model parameters and network structure affect their dynamics~\cite{redner2019,lorenz2007continuous,mcquade2019social}. Although it is valuable to study mathematical models of opinion dynamics, one cannot rely on models alone. To make viable forecasts and sufficiently match empirical data, it is desirable to use parameter values that one obtains from real-life measurements. Unfortunately, it is difficult to directly measure parameters of opinion models from empirical observations in a trustworthy way \cite{bohner2011attitudes,fricker2005experimental}. It is also difficult to justify precise choices of the functional forms of the mathematical terms in such models \cite{franci2019model}. Individuals in a population interact with each other in complex and time-dependent ways, and the opinions of individuals and groups can change drastically from both endogenous evolution and exogenous events. Such multifaceted complexities make it difficult to quantify opinions \cite{kozitsin2021opinion} and measure parameter values in a scientifically rigorous way. Consequently, it is useful to develop approaches from the perspective of inverse problems to infer the parameters of opinion models from data observations. Such efforts can help advance methods of validating opinion models and forecasting future dynamics from past observations.

Prior research on models of interacting agents has employed parameter inference in a variety of settings, including disease spread \cite{fb-published,clemenccon2008stochastic}, election forecasts \cite{Porter_SIAM_review}, opinion dynamics \cite{franci2019model,monti2020}, and dynamics that are governed by distance-based interactions \cite{lu2019nonparametric,liu2022random}.
Such works often formulate an inference problem using a regression framework, in which one seeks to determine optimal parameter values for a model to produce output data that is as close as possible to observed data~\cite{albi2023}.
One can also view this process as a numerical execution of ``inversion'' to tune parameters to characterize the true properties of a dynamical process~\cite{dashti2017bayesian,hinze2008optimization}. 
A fundamental topic that ties closely with such numerical-inversion procedures is whether or not parameters are identifiable from the provided data.  
To do this, it is necessary to relate empirical data to model parameters. 
In the present paper, we examine the following two questions: 
\begin{itemize}
\item[]{(1) What type of data uniquely reconstructs parameters in a model of opinion dynamics?} 
\item[]{(2) How much data does one need for such a reconstruction?}
\end{itemize} 
The answers to these questions depend both on the opinion model itself and on the employed numerical-inversion procedure~\cite{uhlmann2013inverse,kirsch2011introduction}. Researchers have studied these questions in the context of optical tomography \cite{arridge1999optical}, seismology \cite{bube1983one}, geophysics \cite{snieder1999inverse}, and many other applications. In our paper, we examine the above questions for a \emph{bounded-confidence models} (BCM) of opinion dynamics \cite{lorenz2007continuous,bernardo2024}.

In a BCM, the opinions of the entities in a population take continuous values, which perhaps represent a range of opinions from very liberal to very conservative on a one-dimensional political spectrum. 
The interactions between these entities encapsulate the idea of  ``selective exposure'' from psychology \cite{frey1986recent,sears1967selective}. When two or more entities interact, they compromise their opinions by some amount if and only if their current opinions are sufficiently close to each other. One can measure such closeness with a scalar ``confidence bound''. Entities that interact compromise their opinions to some extent if and only if the difference between their current opinions is smaller than the confidence bound; otherwise, in most BCMs, their opinions remain unchanged after an interaction~\cite{lorenz2007continuous,bernardo2024}. 

The idea of a confidence bound appears both in agent-based BCMs and in density-based BCMs \cite{lorenz2007continuous}. Agent-based models and density-based models describe opinion dynamics at different scales, and they serve complementary purposes. Agent-based BCMs characterize fine-grained interactions and are useful for examining the opinion trajectories of a discrete number of agents. They are helpful for obtaining insights into the influence of network architecture on dynamics, such as the qualitative characteristics of steady states (e.g., consensus versus polarization versus fragmentation) and the convergence time to attain a steady state \cite{meng2018opinion}. There is a large body of research on elaborating agent-based BCMs with various features, such as by incorporating heterogeneous confidence bounds \cite{chen2020convergence}, media nodes \cite{brooks2020model}, and other extensions. 
Density-based BCMs take a macroscale perspective and examine the evolution of the opinions of infinitely many agents using population densities \cite{ben2003bifurcations}. 
Density-based BCMs describe the macroscale collective behavior of agents, so they are useful for studying ensemble and average effects.
In some situations, one can derive density-based BCMs as mean-field limits of associated agent-based BCMs \cite{chu2022density}.
Such descriptions arise both in opinion dynamics~\cite{ayi2021mean} and in many other models of the collective behavior of a large number of agents \cite{warren2023}, such as in bird flocking and fish schooling \cite{ha2008particle}.

%%%%

In the present paper, we examine a density-based model of opinion dynamics. Such a model describes the time evolution of an opinion density using a kinetic equation. We seek to develop a mathematically rigorous approach to infer the interaction kernel $\theta$ of our model. \add{During the past decade, there has been considerable theoretical and computational progress in the study of inverse kinetic theory \cite{villani2002review}. 
Key prior work in this area has considered parameter inference in the classical Boltzmann equation \cite{Lai_Boltzmann,li2022determining} and the radiative-transfer equation \cite{hellmuth2022kinetic,CS2,LiSun_2020,BalMonard_time_harmonic}. Inference results in inverse kinetic theory depend heavily on the particular form of a kinetic equation, and it is thus important to examine a variety of models.}

\add{In prior investigations, researchers have examined the inference of interaction kernels in agent-based models using nonparametric methods~\cite{lu2021learning,lu2019nonparametric} and maximum-likelihood methods~\cite{lenti2023}.
Such models {have} finitely many agents and often take the form of a stochastic process or a system of ordinary differential equations.
To obtain a specified accuracy, both the number of data measurements and the computational cost typically increase with the number of agents~\cite{lu2019nonparametric,wang2018inferring,brunton2016discovering}.
In the present paper, we consider a density-based model, which gives a mean-field description when there are infinitely many agents.
Therefore, our inference procedure is in a regime that was not considered in~\cite{lu2019nonparametric,lu2021learning,lenti2023}.
}
%%%%

Our paper proceeds as follows. In section \ref{sec: setup}, we present the mean-field opinion model and set up an inverse problem to infer the interaction kernel between the agents in a population. In section \ref{sec: main-results}, we state and prove two theorems that give theoretical guarantees in the form of sufficient conditions for the data to uniquely and precisely reconstruct the interaction kernel. In section \ref{sec: loss-function}, we propose a numerical method to infer the interaction kernel using a differential-equation-constrained optimization framework. In section \ref{sec: numerics}, we develop an adaptive optimization algorithm to accelerate the convergence of our numerical method. In section \ref{sec: conclusion}, we conclude and discuss future work. 
In Appendix \ref{sec: derive-Lstar}, we derive an adjoint problem, which we use to obtain an explicit formula for an associated Fr\'echet derivative.
In Appendix \ref{sec: derive-M-derivative}, we prove Theorem \ref{thm: M-d}.

%%%%%

\section{Inverse-problem setup} \label{sec: setup}

In this section, we present a mean-field model of opinion dynamics~\cite{ben2003bifurcations} and propose a sensible way to measure data for opinion dynamics.

%%%%

\subsection{A mean-field model of opinion dynamics}

We consider a density-based BCM \cite{ben2003bifurcations,chu2022density} whose governing equation is the Kac-type integro-differential equation 
\begin{equation}\label{eq: pairwise}
	\resizebox{.91\hsize}{!}{$\partial_t{\ftheta}(x,t) = \int_{\Omega\times\Omega}  \!\theta(x_1\!-\!x_2) \ftheta(x_1,t)\ftheta(x_2,t) \!\left[ 2\delta\!\left(x\!-\!\frac{x_1+x_2}{2}\right) - \delta(x\!-\!x_1) - \delta(x\!-\!x_2) \right] dx_1\,dx_2\,,$}
\end{equation}
where $\ftheta(x,t)$ is the probability density of agents with continuous-valued opinion $x \in \Omega$, the set $\Omega\subset\mathbb{R}$ is the space of possible opinions of an agent, $\theta(\cdot)$ is an interaction kernel, and $\delta(\cdot)$ is the Dirac delta function. We use the subscript in $\ftheta$ to indicate explicitly that the solution depends on the interaction kernel $\theta$.
In \eqref{eq: pairwise}, two agents with opinions $x_1$ and $x_2$ interact with each other with a probability that is proportional to $\theta(x_1-x_2)f(x_1,t)f(x_2,t)$. After this interaction, the two agents compromise with each other and change their opinions to their mean opinion $\frac{x_1+x_2}{2}$. This process leads to a ``gain term'' in \eqref{eq: pairwise} at $x=\frac{x_1+x_2}{2}$ from the post-interaction opinions and ``loss terms'' at $x = x_1$ and $x=x_2$ from the pre-interaction opinions. 
One can extend the model \eqref{eq: pairwise} to situations in which the post-interaction opinions have more complicated forms \cite{toscani2006kinetic}, instead of only considering compromises to the mean opinion of two interacting agents. One can also examine generalizations of \eqref{eq: pairwise} that incorporate interactions between three or more entities \cite{chu2022density}.

The interaction kernel $\theta(x_1-x_2)$ encodes the probability that two agents, with opinions $x_1$ and $x_2$, interact with each other. It can take various forms, depending on the specific BCM.
For example, in the classic Deffuant--Weisbuch BCM \cite{deffuant2000mixing}, interacting agents exchange their opinions if their opinion difference is less than a constant $c$. The interaction kernel thus takes the form of the indicator function~\cite{ben2003bifurcations} $\theta(r) = \mathbbm{1}_{(-c,c)}(r)$, which is parameterized by a constant confidence bound $c$.
By contrast, S{\^\i}rbu et al. \cite{sirbu2019algorithmic} examined a BCM that favors edges between nodes whose opinions are very close to each other. They implemented this favoritism using an interaction kernel with a power-law decay. 
The interaction kernel $\theta$ plays a critical role in determining the dynamics of a BCM or other model of collective behavior. For example, Motsch and Tadmor showed that heterophilous interactions promote convergence to consensus in models of interacting agents that adjust to environmental averages~\cite{motsch2014heterophilious}.
To ensure the well-posedness of \eqref{eq: pairwise}, we assume that the interaction kernel is nonnegative \cite{chu2022density}. For simplicity, we also assume that $\theta(r)$ is symmetric around $r = 0$.

A natural question is the following: Given the time series of opinions of all agents, can we infer the opinion-update rule that governs how agents interact and compromise with each other? In the context of the BCM~\eqref{eq: pairwise}, this question amounts to inferring the interaction kernel $\theta$ from observations of the probability density $f(x,t)$. This problem thus falls into the framework of inverse problems of kinetic equations. 

%%%%

\subsection{Accessibility of data}

It is difficult to directly measure people's opinions as continuous values \cite{kozitsin2021opinion}. Instead of assuming that we have a direct measurement of opinions as a function of time, we suppose that the available data takes the form of aggregated values from a distribution of opinions. Each data point is the cumulative probability density up to a value $a$ at time $t$. That is, each data point takes the form
\begin{equation} \label{eq: mf_summary}
    \Mtheta(a,t;f_0) = \int_{-\infty}^{a} f_\theta(x,t) ~dx\,,
\end{equation}
where $\ftheta$ is the opinion-distribution density, which is governed by the mean-field opinion model \eqref{eq: pairwise}, and $f_0$ is its associated initial density.
We use the subscript $\theta$ to emphasize the dependence on the interaction kernel $\theta$.

We refer to $a$ as the \emph{measurement threshold}, which we assume is a known quantity that we are given along with the data. Consider a situation in which people vote in a binary way, such as by choosing ``No'' or ``Yes''. We assume that a vote for ``No'' results from an underlying continuous opinion value that lies in the interval $(-\infty, a]$ and that a vote for ``Yes'' results from a continuous opinion value in the interval $(a, +\infty)$. 
In this example, $\Mtheta$ is the fraction of a population that votes ``No''. The quantity $\Mtheta(a,t;f_0)$ is a function of the measurement threshold $a$, the time $t$, and the initial opinion density $f_0$.
It is impractical to assume that $\Mtheta$ is accessible on the entire domain for all $a$, $t$, and $f_0$. To cope with this reality, we suppose that we only have access to data at certain values of $a$, $t$, and $f_0$.

%%%%%

\section{Theoretical guarantees in two scenarios} \label{sec: main-results}
We consider two specific scenarios: either $\Mtheta$ is measured at a fixed measurement threshold $a$ or it is measured for a fixed initial opinion distribution $f_0$. Both scenarios correspond to certain simplistic real-life situations. For each scenario, we prove that the associated inverse problem is well-defined in the sense that the data uniquely determines the interaction kernel $\theta$.

%%%%%%

\subsection{Data measured at a fixed measurement threshold $a$} \label{sec: single-a-thm}
Suppose that we fix the measurement threshold $a = a_0$ and allow the initial opinion distribution $f_0$ to vary. We define
\begin{equation} 
    m_\theta(t;f_0)=\Mtheta(a_0,t;f_0)\,,
\end{equation}
which is the fraction of opinions that are less than or equal to the threshold $a_0$. This scenario models a situation in which the collected data has a binary nature. This situation occurs frequently in political systems, such as whether the United Kingdom should remain part of the European Union or leave the European Union in the 2016 ``Brexit'' referendum. Another example is a presidential election with two candidates.

We define a functional $\mu_\theta: \mathcal{L}^1(\Omega) \rightarrow \mathbb{R}$ that maps the initial opinion distribution $f_0$ to a real number. This functional,
which is given by
\begin{equation} \label{eq: mu}
    \mu_\theta[f_0] = \partial_t {m}_\theta(t=0;f_0) \,,
\end{equation}
evaluates the rate of change of the ``left-wing'' opinion fraction $m_\theta$ at the initial time $t = 0$, given the initial opinion distribution $f_0$. 
\add{In this scenario, we always fix the measurement threshold $a$ to the value $a_0$.}
One key result of our paper is that the measurement $\mu_\theta$ is sufficient to infer the interaction kernel $\theta$ in Equation~\eqref{eq: pairwise}. In particular, if we know the functional values for the entire function space $\mathcal{L}^1(\Omega)$ (i.e., $\mu_\theta[f_0]$ is available for all $f_0\in\mathcal{L}^1(\Omega))$, then we have sufficient data to uniquely determine $\theta$.
We state the result in Theorem \ref{thm: single-measurement}, which we prove in section \ref{sec: proof}. 

\begin{theorem} \label{thm: single-measurement}
Let $\theta$ be an interaction kernel, and let $\mu_\theta$ be the functional~\eqref{eq: mu}. There is a one-to-one correspondence between $\theta$ and $\mu_\theta$.
\end{theorem}

Theorem \ref{thm: single-measurement} implies that different interaction kernels $\theta_1$ and $\theta_2$ (with $\theta_1 \neq \theta_2$) induce different functionals $\mu_{\theta_1}$ and $\mu_{\theta_2}$. 
Namely, there exists at least one initial opinion distribution $f_0$ such that values of $\mu_{\theta_1}[f_0]$ and $\mu_{\theta_2}[f_0]$ are not equal. This implies that a measurement that includes $\mu_\theta[f_0]$ for all $f_0 \in \mathcal{L}^1(\Omega)$ yields a data set that is sufficient to uniquely reconstruct an interaction kernel. 

At first sight, the required data seems to be rather substantial because we need to evaluate the functional $\mu_\theta[f_0]$ for all initial opinion distributions $f_0$.
However, this strict data requirement is unavoidable to ensure the unique reconstruction of $\theta$. 
The interaction kernel $\theta(r)$ is itself a function, so inferring it requires one to deal with an infinite-dimensional function space.
The observed data needs to be sufficiently abundant to do this. Because the measurement threshold $a$ is fixed at a single value $a_0$, we require flexibility in the initial opinion distribution $f_0$. For example, if the initial opinion distribution $f_0(x)$ is a Dirac delta function, then all nonnegative symmetric kernels yield the same solution $f(x,t) = f_0(x)$. 
In this case, no measurement is able to identify the kernel $\theta$ with this initial opinion distribution.
Another reconstruction failure occurs when the initial opinion distribution $f_0(x)$ is symmetric about $x = a_0$. By symmetry, $m(t;f_0) = 0.5$ for all interaction kernels (i.e., all nonnegative and symmetric functions), so it is not possible to identify the true interaction kernel. 
In the proof of Theorem \ref{thm: single-measurement}, we show that one can slightly relax the stated conditions. Instead of requiring $\mu_\theta[f_0]$ for all $f_0$, we only need $\mu_\theta[f_0]$ for a basis of the $\mathcal{L}^1(\Omega)$ function space.

In practice --- both in reality and in our computations --- we can work with data that consists of a finite number of data points, rather than infinitely many of them. Even with only a limited number of data points, we can still numerically reconstruct the interaction kernel $\theta$. For example, in section~\ref{sec: numerics}, we consider a data set that includes data from 4 different initial conditions. In this example, our numerical inference method is able to reconstruct the interaction kernel up to an accuracy of $10^{-4}$ within 1000 iterations.

%%%%

\subsection{Data measured for a fixed initial opinion distribution $f_0$} \label{sec: single-f0-thm}
In our second scenario, we fix the initial opinion distribution $f_0$ but have data at multiple measurement thresholds. That is, we fix $f_0$ in \eqref{eq: mf_summary} and measure $M_\theta$ for different values of $a$. 
In such a scenario, individuals' opinions about a specified topic take discrete values, which may correspond to satisfaction levels or happiness levels~\cite{helliwell2012world}.
Some surveys also ask participants to rate their views on some scale (e.g., using a Likert scale \cite{likert2021}, which is a common psychometric scale), such as by choosing an integer between $1$ and $10$. 
\add{If one obtains the initial opinion distributions $f_0$ by shifting the same opinion distribution, the second scenario becomes equivalent to the first scenario (see section \ref{sec: single-a-thm}).}

Recall that $M_\theta(a,t;f_0)$ [see \eqref{eq: mf_summary}] indicates the fraction of a population whose opinion has a value of at most $a$.
We define a function $\nu_\theta: \Omega \rightarrow \mathbb{R}$ that depends on the measurement threshold $a$ and $\nu_\theta$. This function takes the form
\begin{equation} \label{eq: nu}
	    \nu_\theta(a) = \partial_t {M}_\theta(a, t = 0; f_0)\,.
\end{equation}

We now present our second main result, which states that the measurement $\nu_{\theta}$ is sufficient to uniquely reconstruct $\theta$.
In particular, if we know the function values $\nu_{\theta}(a)$ for all $a \in \Omega$, we can precisely reconstruct the interaction kernel $\theta$. We state the result in Theorem~\ref{thm: single-initial}, which we prove in section \ref{sec: proof}.

\begin{theorem} \label{thm: single-initial}
Suppose that the interaction kernel $\theta(r)$ is compactly supported on the interval $(-B,B)$ and that the initial opinion distribution $f_0$ that we use to define $\nu_\theta$ in~\eqref{eq: nu} is uniform (i.e., $f_0(x) = \frac{1}{2B}$ if $x\in(-B,B)$ and $f_0(x) = 0$ otherwise). There is then a one-to-one correspondence between $\theta$ and $\nu_\theta$.
\end{theorem}

Theorem \ref{thm: single-initial} implies that for any interaction kernel $\theta_1$ that differs from the true kernel $\theta$, its induced functional $\nu_{\theta_1}$ is also different from $\nu_{\theta}$ at least at one point (i.e., $\nu_{\theta_1}(a_0)\neq\nu_{\theta}(a_0)$ for some $a_0$). Therefore, the data set is sufficient to uniquely identify the true interaction kernel if the data set includes a measurement for all values of $a$. 

Theorem \ref{thm: single-initial} states that 
if we measure opinions at sufficiently finely-grained opinion levels, then a single initial opinion distribution yields enough
data to uniquely reconstruct the interaction kernel. Because the interaction kernel $\theta$ is a function, inferring it precisely requires measurements of different $a$ in a continuous manner, which yields infinitely many data points. As in the scenario in section \ref{sec: single-a-thm}, we only possess a discrete version of the data in practice. In other words, instead of knowing $\Mtheta(a,t;f_0)$ for all $a$ in a continuous interval, we only have a data set $\{M(a,t;f_0):~t\in\mathcal{T},\, a\in\mathcal{A}\}$, where $\mathcal{T}$ and $\mathcal{A}$ are finite sets (see section~\ref{sec: loss-function}). As we enlarge the measurement-threshold set $\mathcal{A}$, we can reconstruct the interaction kernel with better accuracy. 

%%%%%

\subsection{Proofs of Theorems~\ref{thm: single-measurement} and~\ref{thm: single-initial}} \label{sec: proof}
We now prove Theorems~\ref{thm: single-measurement} and~\ref{thm: single-initial}. To ensure mathematical rigor, we first need to justify that the functional $\mu_\theta$ in Equation~\eqref{eq: mu} and the function $\nu_\theta$ in Equation~\eqref{eq: nu} are well-defined. From results in~\cite{chu2022density}, we know that Equation~\eqref{eq: pairwise} is well-posed. In particular, for any nonnegative and symmetric $\theta(r)$ and any density function $f_0(x) \in \mathcal{L}^1(\Omega)$, there exists a unique solution $\ftheta(x,t)$ of Equation~\eqref{eq: pairwise}, with initial opinion distribution $f(x,0) = f_0(x)$, such that $\ftheta(x,t)$ is differentiable with respect to time $t$ and $\int_\Omega\ftheta(x,t)~dx = 1$ for all $t \ge 0$. This implies that (1) the value of the functional $\mu_\theta[f_0]$ is well-defined for all interaction kernels $\theta(r)$ and all initial opinion distributions $f_0 \in \mathcal{L}^1(\Omega)$ and that (2) the value $\nu_\theta(a)$ is well-defined for all interaction kernels $\theta(r)$ and all $a \in \Omega$. 

We now prove Theorem~\ref{thm: single-measurement}, which guarantees that we can uniquely reconstruct an interaction kernel from data 
that we measure at a single measurement threshold.

\begin{proof}[\textbf{Proof of Theorem~\ref{thm: single-measurement}}]
A direct computation yields
\begin{equation} \label{eq: partial-M}	\mu_\theta[f_0]=\partial_t m_\theta(0;f_0) = 2\int_0^{\infty} \theta(y) k(y;f_0)~dy\,,
\end{equation}
where 
\begin{equation} \label{eq: kernel}
	k(y;f_0)= \int_0^{y/2}\! \left[ f_0(a\!-\!y\!+\!x)f_0(a\!+\!x)\!-\!f_0(a\!-\!y\!-\!x)f_0(a\!-\!x)\right]~dx
\end{equation}
is the integral kernel. We choose the initial opinion distribution $f_0$ to be a sum of indicator functions. For any $c$ and $w > 0$, we construct the initial opinion distribution
\begin{equation} \label{eq: f0-indicator}
    f_0(x) = \frac{1}{2w} \left[ \mathbbm{1}_{(a-c-w, a-c)}(x) + \mathbbm{1}_{(a+c-w, a+c)}(x)\right]\,,
\end{equation}
where $\mathbbm{1}_{(l,r)}$ is the indicator function on the interval $(l,r)$. The amplitude $1/(2w)$ ensures that $f_0$ is normalized. For this specific type of $f_0$, we obtain
\begin{equation}\label{triangle}
   k(y;f_0) = \frac{4}{w}h_{2c,w}(y)\,,
\end{equation}
where $h_{2c,w}(y)$ is the triangular hat function that is centered at $y = 2c$ and has a base of width $2w$. That is, 
\begin{equation} \label{eq: h2c}
   h_{2c,w}(y) = \max \left\{1-|(y-2c)/w|\,,0\right\}\,,
\end{equation}
where $\max\{\cdot,\cdot\}$ denotes the larger number of its two arguments.

By considering sufficiently many values of $c$ and $w$,
the set of triangular hat functions $h_{2c,w}(y)$ gives a basis of the $\mathcal{L}^1(\mathbb{R}_{+})$ function space. Therefore, given two functions $\theta_1$ and $\theta_2$, there must exist a pair $c$, $w$ such that
\begin{equation}
    \int_0^\infty \left[ \theta_1(y) - \theta_2(y) \right]h_{2c,w}(y) ~dy \neq 0\,.
\end{equation}
Using Equation~\eqref{eq: partial-M}, we know that $\partial_t m_{\theta_1}(0;f_0) \neq \partial_t {m}_{\theta_2}(0;f_0)$ when $f_0(x)$ takes the form in Equation~\eqref{eq: f0-indicator}. This implies  that $\mu_{\theta_1}[f_0] \neq \mu_{\theta_2}[f_0]$ and thus that $\mu_{\theta_1} \neq \mu_{\theta_2}$. Because we can choose $\theta_1$ and $\theta_2$ arbitrarily, we know that the correspondence between $\theta$ and $\mu_\theta$ is one-to-one.
\end{proof}

%%%%%

We now prove Theorem~\ref{thm: single-initial}, which guarantees that we can uniquely reconstruct an interaction kernel from data 
that we measure for a fixed initial opinion distribution.

\begin{proof}[\textbf{Proof of Theorem~\ref{thm: single-initial}}]
A direct computation yields the Fredholm integral 
\begin{equation} \label{eq: Gint}
\begin{aligned}
    {\partial_{a}}\nu_\theta(a) &= \frac{1}{2B^2}\int_{0}^{B} \theta(y) \, G(a,y)~dy\,,
\end{aligned}
\end{equation}
where
\begin{equation}
    G(a,y) = 4B^2 \left[2 f_0\!\left(a-\frac{y}{2}\right)f_0\!\left(a+\frac{y}{2}\right) - f_0\!\left(a\right)f_0\!\left(a+y\right)- f_0\!\left(a\right)f_0\!\left(a-y\right)\right]
\end{equation}
is the integral kernel. Recall that $f_0$ is a uniform function on $(-B, B)$. A direct computation yields 
\begin{equation} \label{eq: G}
    G(a,\cdot) = \begin{cases}
    \mathbbm{1}_{(B-a,B)}(\cdot)\,, \quad a \in (0,B/2] \\
    \mathbbm{1}_{(B-a,2(B-a))}(\cdot)-\mathbbm{1}_{(2(B-a),B)}(\cdot)\,, \quad a \in (B/2,B)\,.
    \end{cases}
\end{equation}
We claim that $\mathbbm{1}_{(c,B)}(\cdot)$ is a linear combination of $G(a,\cdot)$ with different values of $a$. We prove this claim for the two possible cases: (1) $c \in [B/2, B)$ and (2) $c \in (0, B/2)$.

For case (1), we have
\begin{equation} \label{indicate}
    \mathbbm{1}_{(c,B)}(\cdot)=G(B-c,\cdot) \,\,\, \text{for all}\,\,\, c \in [B/2,B)
\end{equation}
directly from \eqref{eq: G}.
We prove case (2) by induction.
For any $c\in(0,B/2)$, we have
\begin{equation} \label{eq: induction}
    \mathbbm{1}_{(c,B)}(\cdot)= \mathbbm{1}_{(c,2c]}(\cdot) + \mathbbm{1}_{(2c,B)}(\cdot) = G(B-c,\cdot) + 2\mathbbm{1}_{(2c,B)}(\cdot)\,.
\end{equation}
When $c \in [B/4,B/2)$, a direct computation from \eqref{eq: induction} yields
\begin{equation}
    \mathbbm{1}_{(c,B)}(\cdot) = G(B-c,\cdot) + 2G(B-2c,\cdot)\,,
\end{equation}
which implies that $\mathbbm{1}_{(c,B)}(\cdot)$ is a linear combination of $G(a,\cdot)$ when $c \in [B/4,B/2)$. Suppose that $\mathbbm{1}_{(c,B)}(\cdot)$ is a linear combination of $G(a,\cdot)$ for $c \in [B/2^{k},B/2^{k-1})$. From \eqref{eq: induction}, we know that $\mathbbm{1}_{(c,B)}(\cdot)$ is a linear combination of $G(a,\cdot)$ for $c \in [B/2^{k+1},B/2^{k})$. By induction, $\mathbbm{1}_{(c,B)}(\cdot)$ is a linear combination of $G(a,\cdot)$ for $c\in (0,B/2)$.

The above two cases imply that the set $\{G(a,\cdot)\}_a$ is a basis of the function space $\mathcal{L}^1(0,B)$. For any two distinct interaction kernels $\theta_1$ and $\theta_2$, there exists at least one basis function of $\mathcal{L}^1(0,B)$ (e.g., $G(a^*,\cdot)$) such that 
\begin{equation}
    \int_0^{B} \left[\theta_1(y)-\theta_2(y)\right] G(a^*,y)~dy \neq 0\,.
\end{equation}
Equation~\eqref{eq: Gint} implies that $\partial_a \nu_{\theta_1}(a^*) \neq \partial_a \nu_{\theta_2}(a^*)$, which in turn implies that $\nu_{\theta_1} \neq \nu_{\theta_2}$. This concludes the proof.
\end{proof}

We used the same strategy to prove Theorems \ref{thm: single-measurement} and \ref{thm: single-initial}. In this approach, one rewrites the proposed measurement as a Fredholm integral (see Equations \eqref{eq: partial-M} and \eqref{eq: Gint}) and proves that the Fredholm kernel spans the function space as one varies one of the variables. When it does, the reconstruction is unique in the dual space. Researchers use this type of strategy for many inverse problems~\cite{uhlmann2013inverse,kirsch2011introduction}, such as linearized Calder\'on problems and linearized inverse-scattering problems.

In the present paper, we only consider strategies to select either the initial opinion distribution $f_0$ or the measurement threshold $a$ for which the generated data yields a unique inferred interaction kernel $\theta$ in \eqref{eq: pairwise}. However, there are other ways that one can design data measurement to ensure the uniqueness of the inferred interaction kernel $\theta$. For example, one can consider different strategies to generate data at specific times. The uniqueness of the inferred interaction kernel is necessary for the well-posedness of the associated inverse problem (which we introduced in section \ref{sec: setup}) and is also fundamental to the development of numerical inference methods.

%%%%

\subsection{Inference stability}
\add{
Theorem \ref{thm: single-measurement} guarantees that we can uniquely reconstruct an interaction kernel $\theta$ from data that is measured at a fixed measurement threshold. We now discuss the stability of this inference problem. We assume that the derivatives of $\theta$ and $f_0$ are sufficiently smooth (i.e., their derivatives of all orders are continuous and bounded).

To examine the stability of $\mu_\theta[f_0]$ with respect to $\theta$, we need to find a function $F$ such that
\begin{equation}\label{eqn:stability}
	\|\theta_1 - \theta_2\|\leq F(\|\mu_{\theta_1} - \mu_{\theta_2}\|)
\end{equation}
when $\theta$ and $\mu_\theta$ are equipped with proper norms. 
If $F(\cdot)$ is linear and $F(0) = 0$, then the reconstruction of $\theta$ is Lipschitz stable for the selected norms.

We define the infinity norm of $\mu$ by
\begin{equation}\label{def:inf_norm_mu}
	\|\mu_{\theta_1} - \mu_{\theta_2}\| = \max_{c,w}\int (\theta_1 - \theta_2)(y)h_{2c,w}(y)dy
\end{equation}
and equip $\theta$ with the standard Banach-space $L_\infty$ norm. 
Suppose that $\|\theta_1 - \theta_2\|_\infty = \tau > 0$. Because $\theta$ is continuous, we know that there is a point $y_0$ such that $|\theta_1(y_0) - \theta_2(y_0)| = \tau$. 
Additionally, if $\theta$ is differentiable (or smoother), there exists a small neighborhood $(y_0 - \epsilon, y_0 + \epsilon)$ in which $|\theta_1(y) - \theta_2(y)| > \tau/2$ for all $y \in (y_0 - \epsilon, y_0 + \epsilon)$.
The size of $\epsilon$ depends on the smoothness of $\theta$; a smoother $\theta$ yields a larger $\epsilon$. Setting $2c_0 = y_0$ and $w_0 = \epsilon$, we define a function $h_{2c_0,w_0}(y)$ using \eqref{eq: h2c}. We insert $h_{2c_0,w_0}(y)$ into~\eqref{def:inf_norm_mu} to obtain
\begin{equation}
	\|\mu_{\theta-1}-\mu_{\theta_2}\| \geq \int (\theta_1-\theta_2)(y)h_{2c_0,w_0}(y)dy > \tau \cdot \epsilon/2 = \frac{\epsilon}{2}\|\theta_1-\theta_2\|_\infty\,.
\end{equation}
Accordingly, we choose $F(\cdot) = {2}/{\epsilon}$ in \eqref{eqn:stability} and obtain the Lipschitz constant ${2}/{\epsilon}$ in \eqref{eqn:stability}. 
}

\section{A numerical inference method}\label{sec: loss-function}
Our theoretical results in section \ref{sec: main-results} identify forms of data that guarantee a unique reconstruction of the interaction kernel. 
\add{Theorems \ref{thm: single-measurement} and \ref{thm: single-initial} require knowledge of the time derivatives $\partial_t {M}_\theta(a, t = 0; f_0)$ for either all measurement threshold values $a$ or all initial opinion distributions $f_0$. Additionally, the data and the to-be-reconstructed interaction kernel both take a continuous form, so they live in infinite-dimensional spaces.
In practice, however, the time derivatives $\partial_t {M}_\theta$ are often inaccessible and both the data sets and the to-be-reconstructed interaction kernel are finite-dimensional.} In this section, we develop a numerical method to infer the interaction kernel of the mean-field opinion model \eqref{eq: pairwise} in this discrete setting.

Suppose that a data set takes the form
\begin{equation}
    \mathcal{D} = \big\{{M}^\ast(a,t;f_0): a\in \mathcal{A}\,,\, t\in \mathcal{T}\,,\, f_0\in \mathcal{F} \big\}\,,
\end{equation}
where ${M}^\ast(a,t;f_0)$ is the measurement \eqref{eq: mf_summary} that is generated by the true interaction kernel $\theta^*$ (which we specify in our numerical computations) and $\mathcal{A}$, $\mathcal{T}$, and $\mathcal{F}$ are finite collections of discrete values of $a$, $t$, and $f_0$, respectively. We use the notation $|\cdot|$ to represent the cardinality of a set.

%%%%%

\subsection{Loss-function formulation}  \label{sec: loss}
We consider the $\ell_2$ error between simulated data and measured data. This yields a loss function 
\begin{equation} \label{eq: loss}
    \mathrm{L}(\theta) = \frac{1}{|\mathcal{A}||\mathcal{T}||\mathcal{F}|} \sum_{a\in\mathcal{A}}\sum_{t\in\mathcal{T}}\sum_{f_0\in\mathcal{F}} \frac12\left[\Mtheta(a,t;f_0) - {M}^\ast(a,t;f_0) \right]^2 \,.
\end{equation}
In our numerical inference, we seek an interaction kernel $\hat{\theta}$ (i.e., a nonnegative and symmetric function) that minimizes the loss function $\mathrm{L}(\theta)$. That is, we seek to determine
\begin{equation} \label{eq: thetah}
    \hat{\theta} = \text{arg min}_{\theta\in {\Phi}} ~ \mathrm{L}(\theta)\,,
\end{equation}
where ${\Phi} = \{\theta(r)\in\mathcal{L}^\infty: ~\theta(r)\geq0\,, ~\theta(-r) = \theta(r)\}$ is the admissible set. If there are any other restrictions on $\theta$ (such as requiring that $\theta$ is compactly supported on an interval), we place the minimization over an admissible set as constraints to account for them.
\add{To reduce overfitting, one can also add a regularization term of $\theta$ to the loss function in \eqref{eq: loss}.}

There are many strategies to minimize the loss function \eqref{eq: loss}. Roughly speaking, existing methods are either gradient-based or Hessian-based. In theory, Hessian-based methods have higher-order convergence rates (so one may expect them to be faster, in principle) than gradient-based methods, but they require the computation of second-order functional derivatives and are thus computationally prohibitive. 
Therefore, we use a gradient-based minimization algorithm; see section~\ref{sec: algorithm} for details. 
Such an algorithm typically involves an iterative update
\begin{equation} \label{eq: iterations}
    \theta_{n+1 } =\theta_{n} - \alpha_n r_n\,, \quad r_n \approx \partial_\theta \mathrm{L}(\theta)\vert_{\theta=\theta_n}\,,
\end{equation}
where $\partial_\theta \mathrm{L}(\theta)$ is the Fr\'echet derivative with respect to $\theta$. We adjust the step size $\alpha_n$ at each iteration to expedite convergence and ensure that $\theta_{n+1}$ is in the admissible set. The descent direction $r_n$ is a modification of $\partial_\theta \mathrm{L}(\theta)$ that ensures that the update does not leave the admissible set. In other words, it does not violate the nonnegativity constraint and remains symmetric. We compute
\begin{equation} \label{eq: theta-derivative}
    \partial_\theta \mathrm{L}(\theta) = \frac{1}{|\mathcal{A}||\mathcal{T}||\mathcal{F}|} \sum_{a\in\mathcal{A}}\sum_{t\in\mathcal{T}}\sum_{f_0\in\mathcal{F}} \left( \Mtheta(a,t;f_0) - {M}^\ast(a,t;f_0) \right) \partial_\theta \Mtheta(a,t;f_0)\,.
\end{equation}
The core of the computation lies in evaluating the Fr\'echet derivative $\partial_\theta\Mtheta$. This gives a function in the updating direction that lies in the same function space as $\theta(r)$. When the context is clear, we omit the dependence on $a$, $t$, and $f_0$ in our notation and write $\partial_\theta \Mtheta(r)$ as a function of only $r \in \Omega_\theta$. In section \ref{sec: adjoint}, we derive the formula for $\partial_\theta \Mtheta$.

%%%%%

\subsection{Computation of the Fr\'echet derivative} \label{sec: adjoint}
To compute the Fr\'echet derivative in Equation~\eqref{eq: theta-derivative}, we view $M_\theta$ as a functional that maps the function $\theta$ to a real number. The Fr\'echet derivative measures the rate of change of $M_\theta$ when one perturbs the function $\theta$. By using variational calculus, we show that evaluating $\partial_\theta \Mtheta$ amounts to solving two integro-differential
equations with specifically designed initial and final conditions. The former is the forward problem, and the latter is the adjoint problem.
The forward problem is~\eqref{eq: pairwise}, which we rewrite as
\begin{equation} \label{eq: feq}
\left\{
\begin{aligned}
    \partial_t{\ftheta}(x,t) &= \int_{\Omega\times\Omega} \!\!\!\theta(x_1-x_2) \ftheta(x_1,t)\ftheta(x_2,t) F(x,x_1,x_2)~dx_1\,dx_2 \\
    \ftheta(x,0) &= f_0(x)\,,
    \end{aligned}
    \right.
\end{equation}
where $F(x,x_1,x_2)= 2\delta\left(x-\frac{x_1+x_2}{2}\right) - \delta(x-x_1) - \delta(x-x_2)$ and $f_0(x)\in \mathcal{L}^1(\Omega)$ is the initial opinion distribution.
The adjoint problem is
\begin{equation} \label{eq: geq}
    \partial_{\tau} \gtheta(x,\tau) = -\Ltheta^*[\gtheta](x,\tau)\,, \quad \gtheta(x,t) = \mathbbm{1}_{(-\infty,a]}(x)\,,
\end{equation}
where the operator $\mathcal{L}_\theta^\ast$ is the adjoint operator of the integral term in~\eqref{eq: feq}. That is,
\begin{equation} \label{adjoint}
	\Ltheta^*[\gtheta](x,\tau) = \int_{\Omega} 2\ftheta(y,\tau)\theta(x-y) \left[2\gtheta\left({\textstyle\frac{x+y}{2}},\tau\right)-\gtheta(x,\tau) - \gtheta(y,\tau)\right] dy\,.
\end{equation}
In Appendix~\ref{sec: derive-Lstar}, we give a detailed derivation of $\mathcal{L}_{\theta}^*$. We state the formula for the Fr\'echet derivative \eqref{eq: theta-derivative} in Theorem \ref{thm: M-d}, which we prove in
Appendix~\ref{sec: derive-M-derivative}.

\begin{theorem} \label{thm: M-d}
Let $\ftheta$ and $\gtheta$ be solutions of Equations~\eqref{eq: feq} and~\eqref{eq: geq}, respectively. The Fr\'echet derivative in \eqref{eq: theta-derivative} satisfies
\begin{equation} \label{eq: Mtheta-eval}
    {\partial_{\theta}} \Mtheta(r) = \int_{0}^t \!\! \int_{\Omega} 2\ftheta(r+y,\tau)\ftheta(y,\tau)\left[ 2\gtheta\left(\frac{r}{2}+y\right) - \gtheta(r+y) - \gtheta(y)\right]~dy \, d\tau\,,
\end{equation}
with $r\in\Omega_{\theta}$. 
\end{theorem}

Evaluating $\partial_\theta\Mtheta(a,t;f_0)$ involves solving the forward problem~\eqref{eq: feq} for $\ftheta$ and the adjoint problem \eqref{eq: geq} for $\gtheta$, where $a$, $t$, and $f_0$ enter the equations as parameters. The initial opinion distribution $f_0$ is the initial condition of the forward problem, and the interval $(-\infty, a]$ is the final condition of the adjoint problem. We solve both problems on the time interval $[0,t]$. 
The minimization of \eqref{eq: loss} is an example of \emph{differential-equation-constrained optimization} because of its relationship (see \eqref{eq: Mtheta-eval}) between the loss function \eqref{eq: loss} and the forward \eqref{eq: feq} and adjoint \eqref{eq: geq} problems.

%%%%%

\subsection{An optimization algorithm for problems with constraints} \label{sec: algorithm}
In section~\ref{sec: loss}, we formulate the numerical inference of the interaction kernel $\theta$ as a minimization problem of the loss function $\mathrm{L}(\theta)$ [see \eqref{eq: loss}]. 
To minimize the loss function, we adopt a gradient-based method and evaluate the Fr\'echet derivative $\partial_\theta \mathrm{L}(\theta)$ [see \eqref{eq: Mtheta-eval}]. Evaluating $\partial_\theta \mathrm{L}(\theta)$ amounts to repeatedly solving the forward~\eqref{eq: feq} and adjoint~\eqref{eq: geq} problems. In particular, for each iteration in \eqref{eq: iterations}, evaluating $\partial_\theta \mathrm{L}(\theta)$ requires solving $|\mathcal{F}|$ forward problems and $|\mathcal{F}||\mathcal{A}|$ adjoint problems. For each (forward or adjoint) problem, we solve the associated integro-differential equation (\eqref{eq: feq} or \eqref{eq: geq})
for $|\mathcal{T}|$ time steps.

To reduce computational cost, it is important to decrease the number of iterations that we need. We design an algorithm (see Algorithm~\ref{alg: minization}) with an adaptive step size to expedite the convergence of the minimization problem while accommodating the nonnegativity constraint of $\theta$. 
We discretize $\theta(r)$ with a uniform grid and store the discretized kernel as a vector $\Theta$. We use a subscript to denote the solution at a particular iteration. For example, $\Theta_r$ is the solution at the $r$th iteration.

\begin{algorithm}[hbt!]
\caption{An adaptive algorithm to minimize \eqref{eq: loss} with a nonnegativity constraint on the interaction kernel $\theta$} \label{alg: minization}
\begin{algorithmic}[1]
\Statex Input: $\mathcal{T}$, $\mathcal{A}$, $\mathcal{F}$, $\Theta_0$, $\alpha$, $\alpha_{\min}$, $\alpha_{\max}$, $n_\text{max}$
\Statex Output: $\Theta_{n_\text{max}}$
\For{ $n=0,1,2,\ldots,n_\text{max}$ }
        {\For{ $f_0$ in $\mathcal{F}$}
        \State Solve the forward problem~\eqref{eq: feq} for $f(x,t; f_0)$
        \State Solve the adjoint problem~\eqref{eq: geq} with $f(x,t; f_0)$ for all $a\in\mathcal{A}$
        \EndFor
        }
\State Compute $r_n=\partial_\theta \text{L}(\Theta_n)$ using Equations~\eqref{eq: theta-derivative} and~\eqref{eq: Mtheta-eval}
\State $\alpha_n=\max\{\alpha, \alpha/\|r_n\|\}$\,; ~ $\widetilde{\Theta}_{n+1}=\Theta_{n}-\alpha_n r_n$ \label{line: normal1}
\If{ $\min\{\widetilde{\Theta}_{n+1}\}\ge0$ } 
    \State $\alpha = \alpha\times2$\,; ~ $\Theta_{n+1}=\widetilde{\Theta}_{n+1}$ \label{line: double} 
\Else
    \State $\alpha=\max\{\alpha/2, \alpha_\text{min}\}$ \label{line: half}
    \State $r_n(\widetilde{\Theta}_{n+1}<0) = 0$\,; ~ $\Theta_{n+1} = \max\{\widetilde{\Theta}_{n+1},0\}$ \label{line: lessthan}
    \State $\alpha^*_{n}=\min\{\alpha_{\max}, \max\{\alpha, \alpha/\|r_n\|\}\}$ \label{line: normal2}
    \State $\widetilde{\Theta}_{n+1}=\Theta_{n}-\alpha^*_n r_n$
    \If{  $\min\{\widetilde{\Theta}_{n+1}\}\ge0$ }
        \State $\Theta_{n+1}=\widetilde{\Theta}_{n+1}$  
    \EndIf
\EndIf
\EndFor 

\Comment{In line \ref{line: lessthan}, the first assignment $r_n(\widetilde{\Theta}_{n+1}<0) = 0$ is entrywise, so we assign the $i$th entry of $r_n$ to $0$ if the $i$th entry of $\widetilde{\Theta}_{n+1}$ is negative.}
\end{algorithmic}
\end{algorithm}

Algorithm~\ref{alg: minization} is a gradient-based minimization algorithm that uses an adaptive step size both to maintain the nonnegativity constraint and to expedite convergence.
We normalize the step size by the norm of the gradient vector (see lines \ref{line: normal1} and \ref{line: normal2}) to try to avoid getting stuck in a local minimum.
As $\Theta_n$ approaches a local minimum, the gradient-vector norm $\|r_n\|$ becomes closer to $0$. This leads to a larger step size than what one would obtain without normalization and helps the algorithm jump over the local minimum. 
When a forward step satisfies the nonnegativity constraint, we double the step size (see line \ref{line: double}) to speed up the convergence; when a forward step breaks the constraint, we halve the step size (see line \ref{line: half}) and introduce a mechanism to adjust the descent direction to ensure that $\Theta_n \ge 0$ (see line \ref{line: lessthan}).
After fixing the descent direction, we adjust the step size again by using the new gradient norm (see line \ref{line: normal2}).
The new gradient vector $r_n$ may have a fairly small norm, which results in an excessive step size. To avoid abrupt changes in the step size, we impose a lower bound $\alpha_{\min}$ and an upper bound $\alpha_{\max}$ of the step size (see lines \ref{line: half} and \ref{line: normal2}). 
Based on our numerical observations, the two parameters ($\alpha_{\min}$ and $\alpha_{\max}$) are necessary for our algorithm to succeed.

There are other available optimization methods to minimize the loss function \eqref{eq: loss}. One appealing choice is to use a stochastic-gradient-descent (SGD) method \cite{bottou2012stochastic,Jin_2018,Chen_2018} with nonnegativity constraints. 
Because all of the sub-loss functions --- which are given by $\left[\Mtheta(a,t;f_0) - {M}^\ast(a,t;f_0) \right]^2$ for some combination of $a$, $t$, and $f_0$ --- in \eqref{eq: loss} have the same minimizer, we expect to obtain a faster convergence rate with our approach than is typically the case for SGD methods. We do not compare our approach to an SGD approach in this paper, as such a comparison is tangential to our main goals, which are to formulate inverse problems for parameter inference in a BCM and to provide theoretical guarantees for our approach.

%%%%%

\section{Numerical computations}
\label{sec: numerics}
We now do some computations to demonstrate the numerical inference method that we proposed in section \ref{sec: loss-function}. We consider data sets (with, necessarily, finitely many data points) that follow the two scenarios in section \ref{sec: main-results}. With our computations, we demonstrate that our approach from section \ref{sec: loss-function} is able to reconstruct the interaction kernel $\theta$ with good accuracy. We also demonstrate that the reconstruction accuracy increases exponentially as we increase the number of data points.

We consider an opinion space $\Omega = [-1,1]$ and generate data with the true interaction kernel $\theta^*(r) = \mathbbm{1}_{|r| < 0.36}$ of the mean-field opinion model \eqref{eq: pairwise}. We discretize $\Omega$ and $\Omega_{\theta}$ with uniform square grids; each square is of size $d_x \times d_r$, with $d_x = 0.02$ and $d_r = 0.19$. 
We deliberately choose $d_r$ to be much larger than $d_x$. Inferring the interaction kernel using a coarser grid than the one that generates the data helps mitigate overfitting problems.
All of our computations use the same initial point $\Theta_0$; we draw each element of $\Theta_0$ independently from the uniform distribution on $(0,1)$. The iteration parameters in Algorithm~\ref{alg: minization} are $\alpha = 0.01$, $\alpha_{\min} = 0.003$, and $\alpha_{\max} = 0.05$. The set of time steps is $\mathcal{T} = 0.2 \times \{0,1,\ldots,100\}$.

%%%%%%

\subsection{Reconstruction using data from a single measurement threshold}
We fix the measurement-threshold set $\mathcal{A} = \{-1/3\}$ and let the initial-condition set $\mathcal{F}$ (i.e., the set of initial opinion distributions) be collections of uniform probability densities on intervals $(-B,B)$ for several values of $B$. We consider $B = 1$, $B = 0.9$, $B = 0.8$, and $B = 0.7$. These values yield the corresponding uniform probability density functions
\begin{equation}
    \begin{aligned}
    f_0^1(x) &= \frac{1}{2}\mathbbm{1}_{(-1,1)}(x)\,, 
    		\quad f_0^2(x) = \frac{1}{2\times 0.9}\mathbbm{1}_{(-0.9,0.9)}(x)\,, \\
    f_0^3(x) &= \frac{1}{2\times 0.8}\mathbbm{1}_{(-0.8,0.8)}(x)\,, 
    		\quad f_0^4(x) = \frac{1}{2\times 0.7}\mathbbm{1}_{(-0.7,0.7)}(x)\,.
    \end{aligned}
\end{equation}
In our numerical computations, $\mathcal{F}$ is one of the following four sets:
\begin{equation}
    \mathcal{F}=\{f_0^1\}\,, \quad
    \mathcal{F}=\{f_0^1,f_0^2\}\,, \quad 
    \mathcal{F}=\{f_0^1,f_0^2,f_0^3\}\,, \quad
    \mathcal{F}=\{f_0^1,f_0^2,f_0^3,f_0^4\}\,.
\end{equation}

\begin{figure}[ht]
    \centering
    \includegraphics[width=0.45\textwidth]{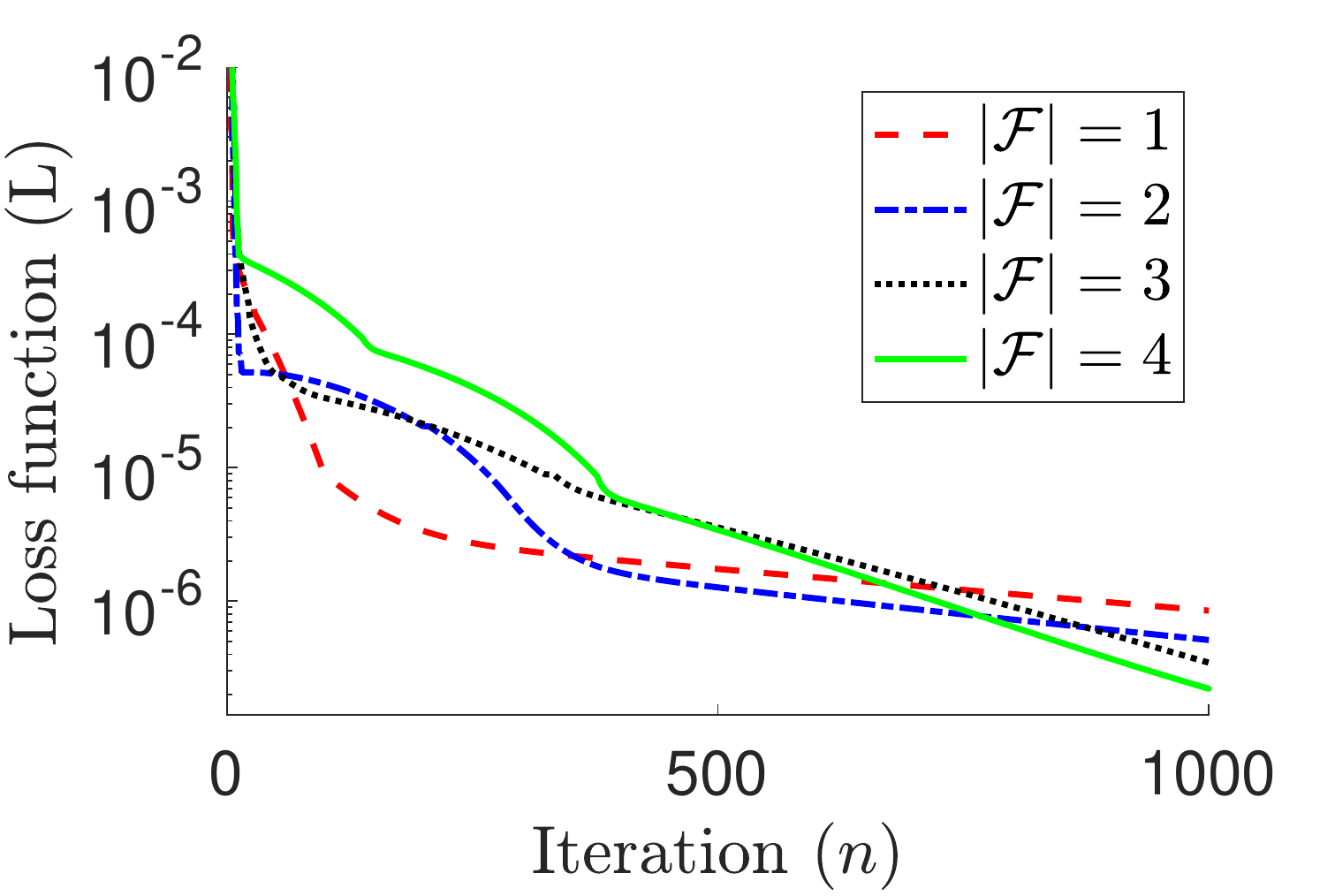} \quad
    \includegraphics[width=0.45\textwidth]{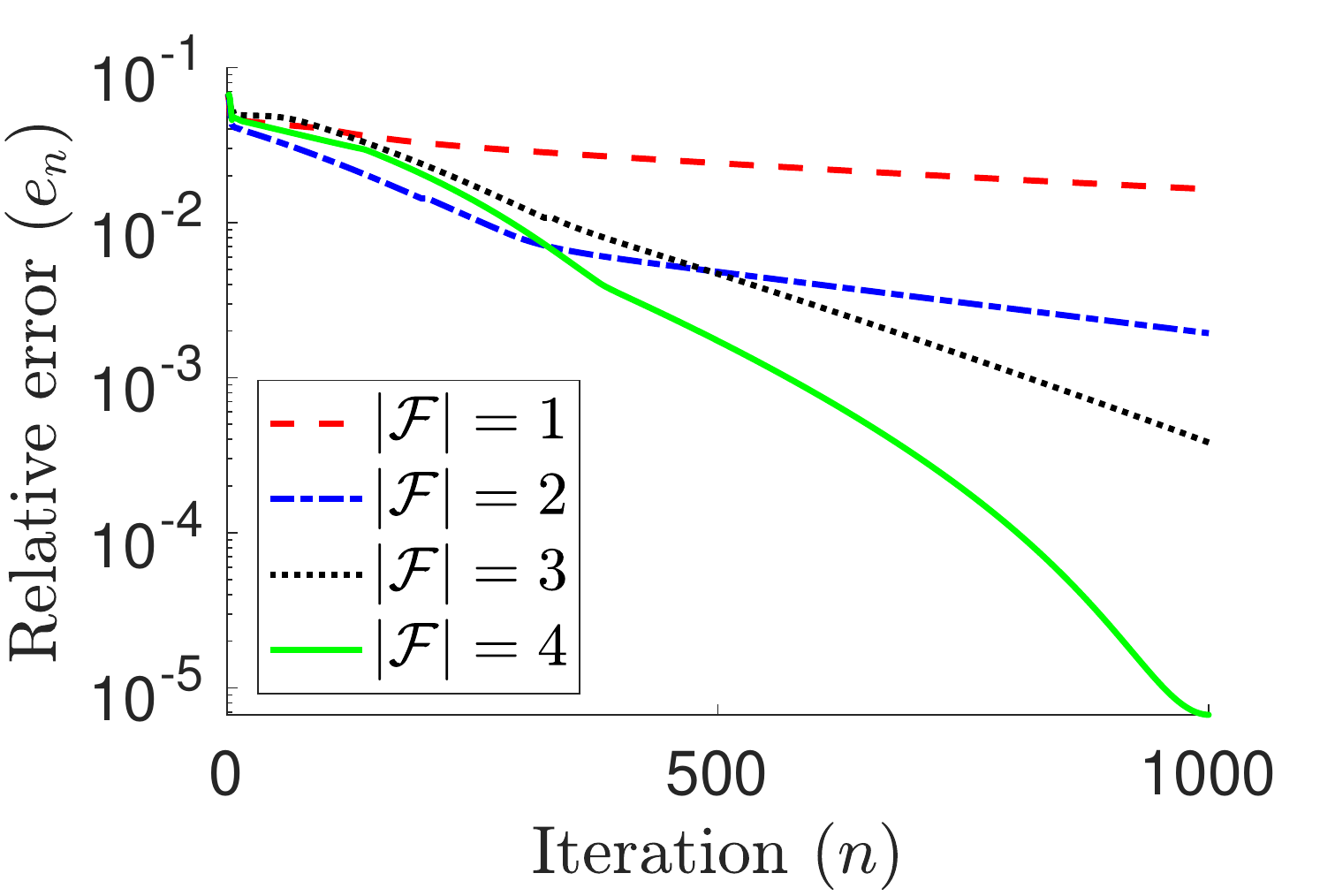}
    \caption{The (left) loss-function value $\mathrm{L}(\Theta_n)$ and (right) relative error $e_n=\|{\Theta}_n-{\Theta}^*\|/\|{\Theta}^*\|$ a function of the iteration $n$ of our optimization algorithm (see Algorithm \ref{alg: minization}).
    We fix the measurement threshold to $a = -1/3$. We perform $n = 1000$ iterations of the algorithm. When the algorithm finishes (i.e., when $n = 1000$), we obtain a relative error of $e_n \approx 1.6 \times 10^{-2}$ for $|\mathcal{F}| = 1$, a relative error of $e_n \approx 1.9 \times 10^{-3}$ for $|\mathcal{F}| = 2$, a relative error of $e_n \approx 3.8 \times 10^{-4}$ for $|\mathcal{F}| = 3$, and a relative error of $e_n \approx  6.7 \times 10^{-6}$ for $|\mathcal{F}| = 4$.
    } 
    \label{fig: single-measurement}
\end{figure}

In Figure~\ref{fig: single-measurement}, we show the decrease of the loss function~\eqref{eq: loss} and the relative error $e_n = {\|\Theta_n-\Theta^*\|}/{\|\Theta^*\|}$ as a function of the iteration $n$ of our optimization algorithm. Early in the optimization process (until about iteration $n = 400$), we observe that the loss function \eqref{eq: loss} decays irregularly. This arises from the adaptive nature of our optimization algorithm. Each iteration uses a new step size and adjusts the gradient-descent direction to preserve nonnegativity. Consequently, the algorithm is unlikely to have a constant convergence rate. In our numerical experiments, the optimization algorithm mostly selects $\alpha_\text{min}$ as the step size after about $400$ iterations. During this phase of the iterative process, the loss function decays exponentially at an approximately constant rate. 

We observe that our optimization algorithm (see Algorithm \ref{alg: minization}) reduces the loss function \eqref{eq: loss} to a value of roughly the same order of magnitude (about $10^{-6}$--$10^{-7}$) after $n = 1000$ iterations for data sets of different sizes ($|\mathcal{F}| = 1,\ldots,4$). In the right panel of Figure \ref{fig: single-measurement}, we observe that the error drops dramatically as we enlarge the data set. 
After $n = 1000$ iterations, even though the loss function attains a similar value (of about $10^{-6}$) for all $|\mathcal{F}| = 1,\ldots,4$, the associated inference of the interaction kernel $\theta$ has different accuracies. With more data points, the loss function \eqref{eq: loss} becomes more effective at measuring the deviation from the true interaction kernel. The computational cost of minimizing the loss function increases as we enlarge the data set. At each iteration, we need to solve $|\mathcal{F}|$ forward problems \eqref{eq: feq} and $|\mathcal{A}|\times|\mathcal{F}|$ (which is equal to $|\mathcal{F}|$ in this example) adjoint problems \eqref{eq: geq}.

%%%%%

\subsection{Reconstruction using data from a single initial opinion distribution}
\begin{figure}[ht] 
    \centering
    \includegraphics[width=0.45\textwidth]{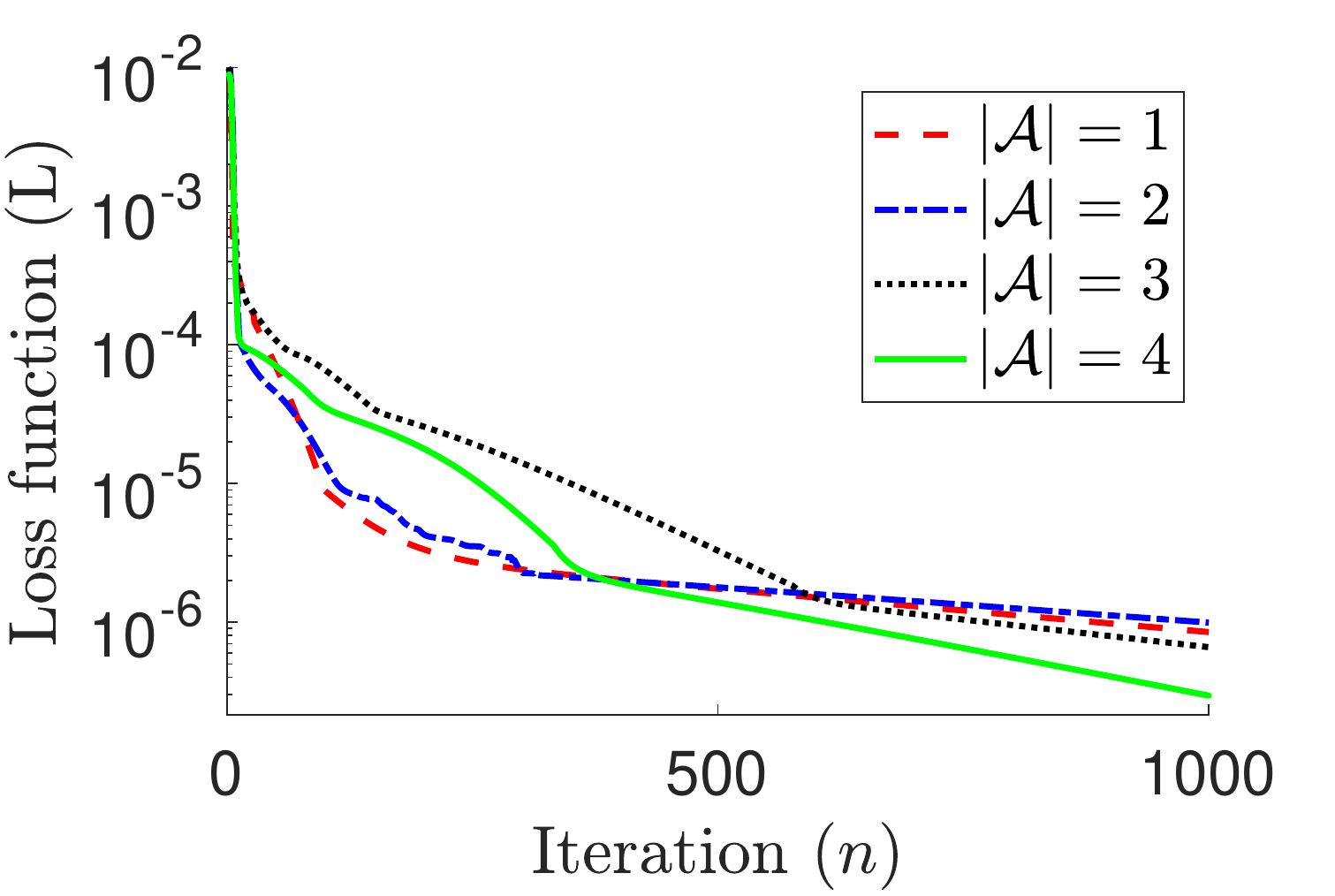} \quad
    \includegraphics[width=0.45\textwidth]{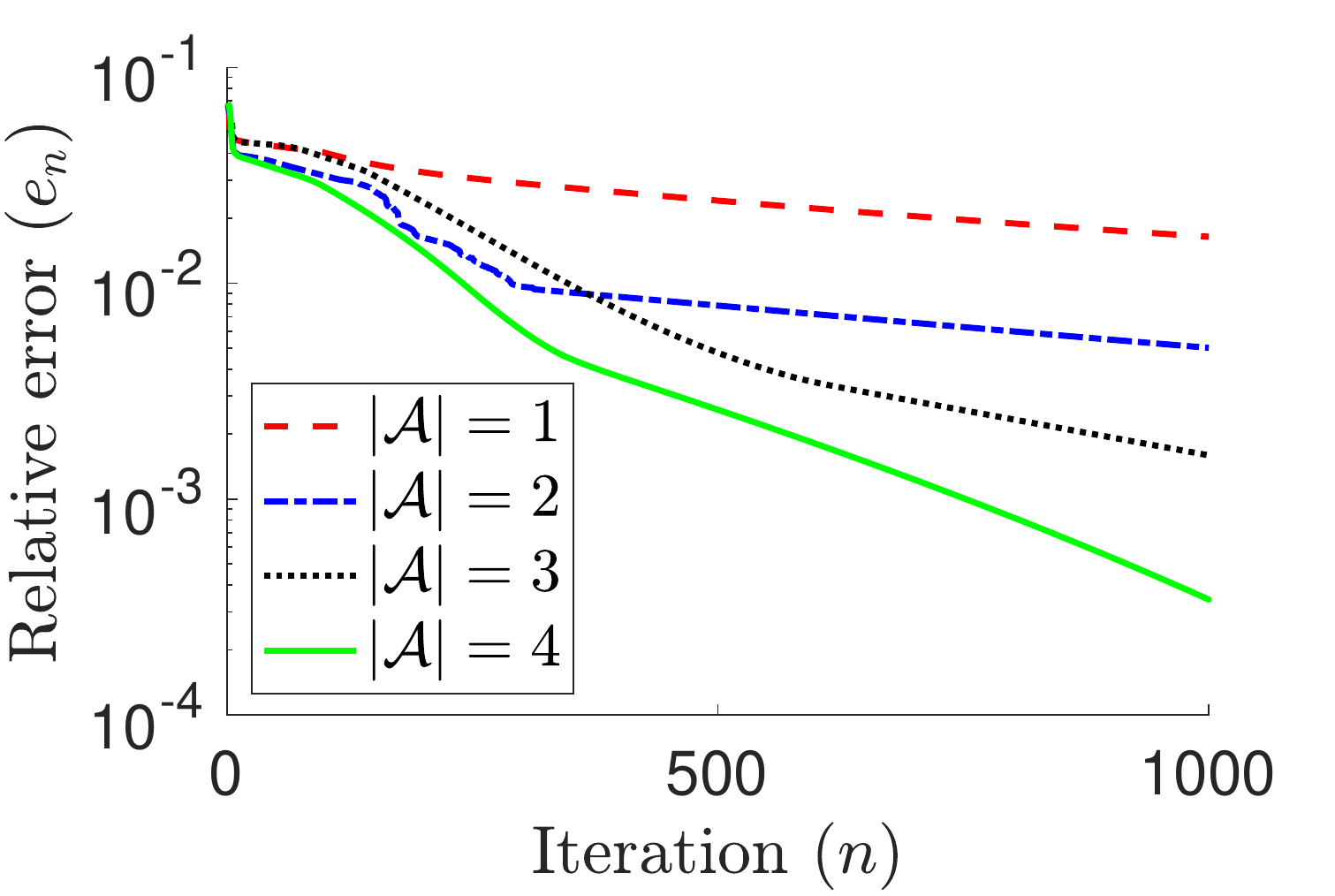}
    \caption{The (left) loss-function value $\mathrm{L}(\Theta_n)$ and (right) relative error $e_n = \|{\Theta}_n-{\Theta}^*\|/\|{\Theta}^*\|$ as a function of the iteration $n$ of our 
    optimization algorithm (see Algorithm \ref{alg: minization}) for a data set that we generate from a single initial opinion distribution. After $n = 1000$ iterations, we obtain a relative error of $e_n \approx 1.6 \times 10^{-2}$ for $|\mathcal{A}| = 1$, a relative error of $e_n \approx 5.0 \times 10^{-3}$ for $|\mathcal{A}| = 2$, a relative error of $e_n \approx 1.6 \times 10^{-3}$ for $|\mathcal{A}| = 3$, and a relative error of $e_n \approx 3.4 \times 10^{-4}$ for $|\mathcal{A}| = 4$.
    }
    \label{fig: single-initial}
\end{figure}

We now fix the initial opinion distribution $f_0(x) = \mathbbm{1}_{(-1,1)}(x)$ and vary the measurement-threshold set $\mathcal{A}$. We take $\mathcal{A}=\{a_0-1,2a_0-1,\ldots, |\mathcal{A}|a_0-1\}$, with $a_0 = 1/(3|\mathcal{A}|)$, and we consider $|\mathcal{A}| = 1,\ldots,4$.
In Figure~\ref{fig: single-initial}, we show the loss-function values and the relative errors as functions of the number of iterations of our optimization algorithm.

In the left panel of Figure~\ref{fig: single-initial}, the value of the loss function \eqref{eq: loss} decreases to about $10^{-6}$ after $n = 1000$ iterations for all data sets (i.e., for all $|\mathcal{A}| = 1,\ldots,4$). Additionally, as in Figure \ref{fig: single-measurement}, we again observe that the loss function decays irregularly for an initial set of iterations (until about $n = 550$) before decaying exponentially (for $ n \ge 550$).
After $n = 1000$ iterations, the loss function decreases to a value of the same order of magnitude for all 4 data sets. However, the relative error between the inferred and true interaction kernels differs across the 4 data sets. As we enlarge the size of the measurement-threshold set $\mathcal{A}$, the relative error decreases dramatically and the loss function becomes more effective at measuring the deviation from the true interaction kernel. 
At each iteration, we need to solve 1 forward problem \eqref{eq: feq} (because we fix the initial opinion distribution $f_0$) and $|\mathcal{A}|$ adjoint problems \eqref{eq: geq}.

%%%

\subsection{Comparing the two scenarios}
\begin{figure}[htp]
    \centering
    \includegraphics[width=0.75\textwidth]{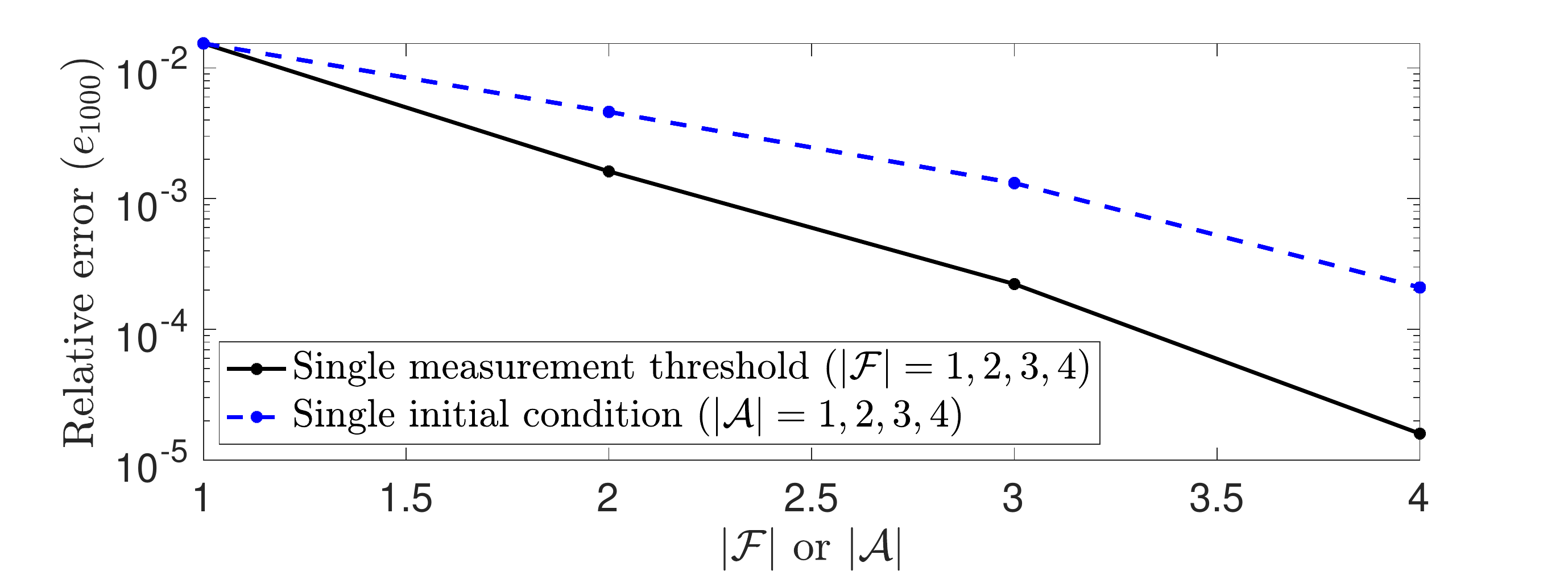}
    \caption{A comparison of the sizes $\|\hat{\Theta} - \Theta\|/\|\Theta\|$ of the relative error between the inferred interaction kernel and the true interaction kernel for our two scenarios. We plot the relative error from data that we measure at a fixed measurement threshold but with different numbers of initial opinion distributions ($|\mathcal{F}| = 1,\ldots,4$) as a solid black curve.
    We plot the relative error from data that we measure for a fixed initial opinion distribution but with different numbers of cumulative thresholds ($|\mathcal{A}| = 1,\ldots,4$) as a dashed blue curve.
    }
    \label{fig: comparison}
\end{figure}

By comparing Figures~\ref{fig: single-measurement} and \ref{fig: single-initial}, we observe that when we terminate the optimization process, the loss-function values in both scenarios are about $10^{-6}$ but that the relative error between the inferred interaction kernel and the true interaction kernel is much smaller for $|\mathcal{F}| = 4$ than for $|\mathcal{A}| = 4$. 
In Figure~\ref{fig: comparison}, we plot the relative error after $n = 1000$ iterations for different values of $|\mathcal{F}|$ and $|\mathcal{A}|$. We observe that the relative error decays faster as we increase $|\mathcal{F}|$ than it does as we increase $|\mathcal{A}|$.

Importantly, our comparison between the two kernel-reconstruction scenarios does not imply that including data from multiple initial opinion distributions is more efficient than including data from multiple measurement thresholds. The performance of our optimization algorithm also depends on the initial opinion distributions and the measurement points. When two initial opinion distributions are too similar, one typically expects them to yield similar dynamics and in turn to yield similar output data (although this need not be the case for chaotic dynamics), which thus may do little or nothing to improve the performance of our optimization algorithm and parameter inference. 
An analogous situation occurs when two measurement points are too close to each other. It is an important goal for future work to develop techniques to assess choices of the initial opinion distribution $f_0$ and measurement threshold $a$ before applying an optimization algorithm to infer an interaction kernel.

%%%%%

\section{Conclusions and discussion} \label{sec: conclusion}
In the mathematical modeling of opinion dynamics, it is typically difficult (or even impossible) to directly observe or measure parameter values or the functional forms of interactions. Therefore, it is important to develop methods to infer unknown parameters (either constants or functions) from empirical opinion data. To explicitly perform such a procedure, we formulated and examined an inverse problem using a mean-field bounded-confidence model (BCM) of opinion dynamics. We inferred the mean-field BCM's interaction kernel, which is a function that encodes how two agents interact and compromise their opinions. We examined this procedure both theoretically and numerically.

In our inverse problems, we considered two types of data sets: one with a fixed initial opinion distribution and the other with a fixed measurement threshold. For both scenarios, 
we proved that the given data has sufficient information to uniquely identify the interaction kernel $\theta$ (i.e., that the associated inverse problem is well-posed).
We then developed a numerical inference strategy that employs a differential-equation-constrained optimization framework and seeks parameter values that produce simulated data that best matches the given empirical data. We formulated a gradient-based algorithm to execute the optimization.
In this algorithm, one computes a Fr\'echet derivative with respect to the interaction kernel $\theta$ and repeatedly solves a set of forward and adjoint problems.
Our numerical results showcased our well-posedness results for both scenarios.

The perspective of inverse problems is promising for the study of opinion dynamics. In the present paper, we examined inverse problems that are associated with a density-based BCM. It is also worthwhile to formulate and analyze inverse problems for agent-based models of opinion dynamics, such as for inferring the discordance function in agent-based BCMs on hypergraphs \cite{hickok2022bounded} or inferring waiting-time distributions in a non-Markovian opinion model on temporal networks \cite{chu2022non}. We expect that it will be valuable to study such inverse problems to validate opinion models and concretely connect them to real-life phenomena. It is also desirable to investigate a variety of approaches for the numerical inference strategy. We employed a simple gradient-based method, and approaches such as Hessian-based methods and interior-point methods may yield computational benefits.

%%%%%%

\appendix

\section{Perturbed equations and adjoint operators} \label{sec: derive-Lstar}
In this appendix, we examine the perturbed dynamics of the forward problem \eqref{eq: feq} and derive the adjoint operator $\Ltheta^*$ [see \eqref{adjoint}]. 

Let $f_{\theta+\thetat}(x,t)$ be the solution of the forward problem~\eqref{eq: feq} with the interaction kernel $\theta + \thetat$, and let $\ft = f_{\theta+\thetat} - f_\theta$. Differentiating $\ft$ with respect to time yields
\begin{equation}\label{eqn:ft}
    \partial_t \ft(x,t) = \Ltheta[\ft] + \Stheta[\thetat]\,, \quad \ft(x,0) = 0\,,
\end{equation}
where
\begin{equation} 
    \begin{aligned}
        \Ltheta[\ft](x,t) &= \int_{\Omega\times\Omega}\!\!\!\! 2\ft(x_1,t)\ftheta(x_2,t)\theta(x_1-x_2)F(x,x_1,x_2)~dx_1\,dx_2\,, \\
           \label{eq: Stheta} 
        \Stheta[\thetat](x,t) &= \int_{\Omega\times\Omega} \!\!\!\! \thetat(x_1-x_2)\ftheta(x_1,t)\ftheta(x_2,t)F(x,x_1,x_2) ~dx_1\,dx_2\,.
    \end{aligned}
\end{equation}
With a direct computation, we see that the adjoint operator $\Ltheta^\ast$ satisfies
\begin{equation}
\begin{aligned}
    \Ltheta^*[g](x,t) &= \int_{\Omega\times\Omega} \!\!\!\!
    2g(x_1,t)\theta(x-x_2)f_\theta(x_2,t)F(x_1,x,x_2)~dx_1\,dx_2 \\
    &= \int_{\Omega} 2\ftheta(y,t)\theta(x-y) \left[2g\left(\frac{x+y}{2},t\right)-g(x,t)-g(y,t)\right]~dy\,.
\end{aligned}
\end{equation}
Let $\gtheta$ be the solution of the adjoint problem 
\begin{equation}\label{eqn:g}
    \partial_t \gtheta(x,t) = -\Ltheta^*[\gtheta](x,t)\,, \quad \gtheta(x,T) = \mathbbm{1}_{(-\infty,a]}(x)\,.
\end{equation}
A direct computation from \eqref{eqn:ft} and \eqref{eqn:g} yields
\begin{equation} \label{thisequation}
    \partial_t(\ft \gtheta) = \Ltheta[\ft]\gtheta + \Stheta[\thetat]\gtheta - \Ltheta^*[\gtheta]\ft \, .
\end{equation}
We integrate both sides of \eqref{thisequation} in both time and space. 
Using the initial opinion distribution (i.e., initial condition) of \eqref{eqn:ft} and the final condition of \eqref{eqn:g}, the left-hand side of \eqref{thisequation} becomes
\begin{equation}
    \int_0^T \!\! \int_\Omega \partial_t(\ft \gtheta)(x,t) ~dx\,dt = \int_{-\infty}^a \ft(x,T) ~dx= \widetilde{M}(T)\,,
\end{equation}
where $\widetilde{M}=M_{\theta+\thetat}-M_{\theta}$ and ${M}_\theta$ is defined in \eqref{eq: mf_summary}.
After integration, the first and last terms of \eqref{thisequation} on the right-hand side cancel each other. Consequently, 
\begin{equation}
\begin{aligned}
    \widetilde{M}(T) &= \int_0^T\!\!\int_\Omega \Stheta[\thetat](x,t)\gtheta(x,t) ~dx\,dt\,.
\end{aligned}
\end{equation}

We conclude the derivations in this appendix with the following lemma.

\begin{lemma} \label{thm: M-derivative}
Let $\ftheta$ and $f_{\theta + \tilde{\theta}}$ be solutions of Equations~\eqref{eq: feq} with interaction kernels $\theta$ and $\theta+\tilde{\theta}$, respectively. Let
$\gtheta$ be the solution of \eqref{eq: geq}. We then have
\begin{equation} \label{eq: Mtilde}
	  M_{\theta+\widetilde{\theta}}(a,t;f_0) - M_{\theta}(a,t;f_0) = \int_0^t\!\!\int_{\Omega} \Stheta[\thetat](x,\tau)\gtheta(x,\tau) ~dx\,d\tau\,,
\end{equation}
where $M_{\theta+\thetat}$ and $M_{\theta}$ are defined in \eqref{eq: mf_summary} using $f_{\theta+\thetat}$ and $f_{\theta}$, respectively, and $\Stheta$ is defined in \eqref{eq: Stheta}.
\end{lemma}

%%%%%%%

\section{Proof of Theorem~\ref{thm: M-d}} \label{sec: derive-M-derivative}
In this appendix, we prove Theorem~\ref{thm: M-d}.

\begin{proof}
Recall Lemma \ref{thm: M-derivative} and the definition of $\Stheta$ in \eqref{eq: Stheta}. We compute
\begin{align}  \label{eq: difference}
    \!\!M_{\theta+\widetilde{\theta}} - M_{\theta}
    &= \int_0^t\!\!\int_\Omega \!\! \Stheta[\thetat](x,\tau)\gtheta(x,\tau) ~dx\,d\tau  \\
    &= \int_{0}^t \!\!\int_{\Omega_\theta\times\Omega\times\Omega} \!\! \thetat(r)\gtheta(x,\tau)\ftheta(r+y,\tau)\ftheta(y,\tau)F(x,r+y,y) ~dx\,dy\,dr\,d\tau \notag \\
    &\, = \int_{0}^t\!\!\int_{\Omega_\theta\times\Omega}\!\! 2\thetat(r)\ftheta(r+y,\tau)\ftheta(y,\tau) \notag \\
    & \qquad  \qquad  \quad \times \left[2\gtheta\left(\frac{r}{2}+y,\tau\right)- \gtheta\left(r+y,\tau\right)- \gtheta(y,\tau)\right]~dy\,dr\,d\tau\,, \notag
\end{align}
where we have omitted writing the dependence on $a$, $t$, and $f_0$.
Equation \eqref{eq: difference} yields the Fr\'echet derivative
\begin{equation}
    {\partial_{\theta}} \Mtheta(r) = \int_{0}^t\!\!\int_{\Omega} 2\ftheta(r+y,\tau)\ftheta(y,\tau)\!\left[2\gtheta\left(\frac{r}{2}+y,\tau\right)- \gtheta\left(r+y,\tau\right)- \gtheta(y,\tau)\right]~dy\,d\tau\,.
\end{equation}

\quad
\end{proof}

%%%%%

\section*{Acknowledgements}
QL was supported in part by the National Science Foundation (through the grants NSF-CAREER-1750488 and NSF-DMS-2023239). MAP was funded by the National Science Foundation (grant 1922952) through the Algorithms for Threat Detection (ATD) program. WC was supported by the Wisconsin Alumni Research Foundation for her visit to QL at UW-Madison, where the research was initiated.

%%%%%

\bibliographystyle{siamplain}
\bibliography{bibfile}

%%%%%%%%%%%%%%%

%%%%%%%%%%%%%%%%%

\end{document}